\documentclass[preprint,showpacs,preprintnumbers,amsmath,amssymb]{revtex4}
\usepackage{graphicx}

\begin{document}

\thispagestyle{empty}

\title{Magnetic materials and the problem of  thermal
Casimir force}

\author{B.~Geyer,${}^{1}$
G.~L.~Klimchitskaya,${}^{1,2}$ and
V.~M.~Mostepanenko${}^{1,3}$}

\affiliation{${}^1$Institute for Theoretical
Physics, Leipzig University, Postfach 100920,
D-04009, Leipzig, Germany \\
${}^2${North-West Technical University,
Millionnaya St. 5, St.Petersburg,
191065, Russia}\\
${}^3${Noncommercial Partnership ``Scientific Instruments'',
Tverskaya St. 11, Moscow,
103905, Russia}
}

\begin{abstract}
We investigate the thermal Casimir interaction between two
magnetodielectric plates made of real materials.
On the basis of the Lifshitz theory, it is shown that
for diamagnets and for paramagnets in the broad sense (with exception
of ferromagnets) the magnetic properties do not influence the
magnitude of the Casimir force. For ferromagnets, taking into account
the realistic dependence of magnetic permeability on frequency, we conclude
that the impact of magnetic properties on the Casimir interaction arises
entirely from the contribution of the zero-frequency term in the
Lifshitz formula. The computations of the Casimir free energy and
pressure are performed for the configurations of two plates made of
ferromagnetic metals (Co and Fe), for one plate made of ferromagnetic
metal and the other of nonmagnetic metal (Au), for two ferromagnetic
dielectric plates (on the basis of polystyrene), and for a ferromagnetic
dielectric plate near a nonmagnetic metal plate. The dielectric
permittivity of metals is described using both the Drude and the plasma
model approaches. It is shown that the Casimir repulsion through the
vacuum gap can be realized in the configuration of a ferromagnetic
dielectric plate near a nonmagnetic metal plate described by the plasma
model.  In all cases considered, the respective analytical results in the
asymptotic limit of large separations between the plates are obtained.
The impact of the magnetic phase transition through the Curie
temperature on the Casimir interaction is considered. In conclusion, we
propose several experiments allowing to determine whether the magnetic
properties really influence the Casimir interaction and to
independently verify the Drude and plasma model approaches to the thermal
Casimir force.

\end{abstract}
\maketitle

\section{Introduction}

At present physical phenomena caused by the zero-point
oscillations of quantized fields attract much experimental
and theoretical attention. One of the most prospective
subjects in this area is the Casimir effect \cite{1}, i.e.,
the attractive force acting between two neutral parallel
ideal metal plates in vacuum arising due to the existence
of zero-point oscillations of the electromagnetic field
and thermal photons. The Casimir force is a version of the
van der Waals force in the case when the separation distances
between the interacting bodies are large enough so that the
relativistic effects contribute essentially. Lifshitz
\cite{2,3} developed the general theory of the van der Waals
and Casimir forces for the case of two dielectric
semispaces separated with a gap of width $a$. The material of
semispaces was described by the frequency-dependent dielectric
permittivity $\varepsilon(\omega)$. In recent years several
reviews \cite{4,5,6,7} and books \cite{8,9,10,11} on different
aspects of the Casimir effect have been published. A new stage
in the measurement of the Casimir force was started by the
two experiments \cite{12,13}. During the {last} few years
significant progress has been made both in the measurement
of the Casimir force and in the development of new calculational
methods applicable to nontrivial geometries and taking into
account real material properties of the interacting bodies.
This progress is reflected in the monograph \cite{14}.

The {seminal} paper by Casimir \cite{1} treated the configuration
of two parallel ideal metal plates which do not posses magnetic
properties and found that the force is always attractive.
The possibility to obtain the repulsive Casimir force has
agitated scientists for several decades. It is of high promise
for solving the problems of stiction and friction in micro-
and nanoelectromechanical devices \cite{15}.
Boyer \cite{16,17} was the first who considered configurations
of an ideal metal spherical shell and of two parallel plates one of
which is made of an ideal metal and another one is infinitely
permeable. In both cases the Casimir force was shown to be
repulsive. The latter configuration which is better adapted for
possible applications in microdevices was often discussed in the
literature as an {\it unusual, hybrid} or {\it mixed} pair of
plates \cite{18,19,20}.

The investigation of the influence of magnetic properties
on the Casimir force in the case of real materials requires the
generalization of the Lifshitz theory for magnetodielectric
media possessing  frequency-dependent dielectric permittivity
$\varepsilon(\omega)$ and magnetic permeability $\mu(\omega)$.
Such generalization was performed by Richmond and Ninham \cite{21}
and later formulated for an arbitrary number of plane parallel
layers of magnetodielectrics \cite{22,23}. As was remarked in the
familiar review \cite{24}, in most of cases the contribution of
magnetic properties of the bodies into the van der Waals
interaction is very small. It was mentioned also that in some
cases, for example, for polarizable particles with both electric
and magnetic polarizabilities \cite{25}, the inclusion of magnetic
properties may be interesting. This was confirmed in the
investigation of the impact of  magnetic properties of
both atoms and material of the
wall on atom-wall interactions including
the case of multiple walls \cite{22,26,27}.

Calculation of the influence of magnetic properties of
plate materials on the Casimir interation between two
magnetodielectric plates was performed in Ref.~\cite{28} using
the approximation of frequency-independent $\varepsilon$ and
$\mu$. Repulsive forces were found in a wide range of
parameters, and the importance of this phenomenon for
experimental study and for nanomachinery applications was
noted. It was shown \cite{29}, however, that for real materials
$\mu$ is nearly equal to unity in the range of frequencies
which gives major contribution to the Casimir force.
As a result, the magnitude of $\mu$ is always far away from
the values needed to achieve the Casimir repulsion \cite{29}.

In this connection, Ref.~\cite{23} reconsidered this problem
for the configuration of one metal and one magnetodielectric
plate taking into account dependences of $\varepsilon$ and
$\mu$ on the frequency.  In so doing the metal (Au) was described
by the Drude dielectric permittivity and the permittivity and
permeability of a magnetodielectric was described by a
simplified model of  the Drude-Lorentz type. It was shown that
at zero temperature there is a repulsive regime, but only at
large separations of about $15\,\mu$m. At nonzero temperature
the Casimir force was found to be always attractive.
It should be taken into account, however, that at room temperature
the theoretical description of the Casimir force by means of the
Lifshitz theory combined with the Drude model is experimentally
excluded at high confidence level \cite{14,30,31}.
Some authors \cite{31a}, however, called for the reanalysis of
electrostatic calibrations in previous experiments on the Casimir
force basing on their measurements with by a factor of 200 larger
sphere radius. This call initiated a discussion in the
literature \cite{31b,31c}.
Further study may be needed here before the situation will
become well understood.
 Because of this,
it is worthwhile to analyze the problem by using different
approaches to the theory of thermal Casimir force suggested in
the literature with allowance made for all existing types
of magnetodielectric materials.

In this paper we investigate the thermal Casimir force between
magnetodielectric plates with different magnetic properties,
and also between a magnetodielectric and a metal plate.
As a magnetodielectric, both diamagnetic and paramagnetic
materials of the plate are considered taking into account
realistic dependence of their $\varepsilon$ and $\mu$ on the
frequency \cite{32,33,34}. It is shown that for all
diamagnets and for paramagnets in the broad sense (with
exclusion of only ferromagnets) the influence of magnetic
properties of plate material on the thermal Casimir force is
negligibly small. This confirms the conclusion made in Ref.~\cite{24}.
Special attention is paid to the case of Casimir plates made of
ferromagnetic materials. From the analysis of frequency dependence
of magnetic susceptibilities of ferromagnets, we arrive to the
conclusion that magnetic properties can influence the thermal
Casimir force only through the contribution of the zero-frequency
term of the Lifshitz formula.

For two similar plates made of ferromagnetic metal the influence
of magnetic properties on the magnitude of the Casimir force
strongly depends on the model of dielectric permittivity used.
Below we show that if $\varepsilon(\omega)$ is represented within
the Drude model approach \cite{35}, the magnitude of the Casimir
force in the high-temperature limit may increase in two times
owing to the account of magnetic properties. If the plasma model
approach \cite{36,37} is used, the magnitude of the Casimir
force at high temperature computed with account of magnetic
properties may be even smaller than in the case when the magnetic
properties are disregarded.
For two similar plates made of ferromagnetic dielectric
the thermal Casimir force at high temperature is shown to be by
a factor of 3 larger owing to the account of magnetic properties.

The role of magnetic properties in the interaction of a ferromagnetic
plate with a nonmagnetic metal plate also strongly depends on the
model of a metal used. We demonstrate that if the Drude model is
used to describe the dielectric properties of two metal plates
one of which is ferromagnetic and the other is nonmagnetic,
the thermal Casimir force is the same as for two nonmagnetic
plates. If, however, both ferromagnetic and nonmagnetic metal plates
are described by the plasma model, the inclusion of magnetic
properties into the Lifshitz theory leads to a decrease in the
magnitude of the Casimir force. For a ferromagnetic dielectric plate
interacting with a nonmagnetic metal plate described by the Drude
model we show that the magnetic properties do not influence the
Casimir force. If the nonmagnetic metal is described be the plasma
model, we find that the account of magnetic properties of
ferromagnetic dielectric leads to a decrease of force magnitude
and may even reverse its sign by changing attraction  for repulsion.
The use of different approaches to the description of dielectric
properties of metal is also shown to influence the behavior of the
Casimir force as a function of temperature in the vicinity of
the Curie temperature of the ferromagnet.

On the basis of the above listed results we propose several
experiments on the measurement of the Casimir force which should be
capable to determine whether or not the magnetic properties
influence the force magnitude and which model of the dielectric
permittivity of metal is experimentally consistent.

The structure of the paper is the following. In. Sec.~II we briefly
introduce the Lifshitz formulas for two dissimilar magnetodielectric
semispaces
and provide necessary information for the dielectric permittivity
and magnetic permeability as functions of frequency.
Section~III is devoted to computations of the Casimir free energy
per unit area and pressure as functions of separation in the
configurations of two thick parallel plates made of ferromagnetic metals.
Some analytic results are also provided.
 In Sec.~IV the configuration
of a ferromagnetic metal plate near a nonmagnetic metal plate
is considered and the Casimir free energy
and pressure are calculated.
In Sec.~V similar computations  are performed for the configurations
where ferromagnetic metal plates are replaced with
ferromagnetic dielectric plates. The dependence of the Casimir force
on the temperature in the vicinity of Curie temperature is
considered in Sec.~VI. Here we show that the behavior of the Casimir
force during the phase transition from the ferromagnetic to
paramagnetic (in a narrow sense) state also critically depends on
the model of $\varepsilon$ of metal plates. In Sec.~VII we present
our conclusions and discussion. Specifically, we suggest a few experiments
which could confirm or exclude the influence of magnetic properties
of plate materials on the Casimir force and help to make a choice
between different approaches to the theoretical description of the
thermal Casimir force.

\section{The Lifshitz formula and real material properties of
magnetodielectrics}

We consider the configuration of two thick dissimilar magnetodielectric
plates (semispaces) separated by a gap of width $a$ at temperature $T$
in thermal equilibrium with the environment. Then, assuming linear
relations between the electric field and electric displacement and
magnetic field and magnetic induction, i.e.
$\mbox{\boldmath$D$}=\varepsilon\mbox{\boldmath$E$}$,
$\mbox{\boldmath$B$}=\mu\mbox{\boldmath$H$}$,
 the Casimir free energy
per unit area of the plates is given by \cite{14,21,22,23,24}
\begin{eqnarray}
{\cal F}(a,T)&=&\frac{k_BT}{2\pi}\sum\limits_{l=0}^{\infty}
{\vphantom{\sum}}^{\prime}\int_0^{\infty}k_{\bot}dk_{\bot}
\left\{\ln\left[1-r_{\rm TM}^{(1)}(i\xi_l,k_{\bot})
r_{\rm TM}^{(2)}(i\xi_l,k_{\bot})\,e^{-2aq_l}\right]\right.
\nonumber \\
&&+\left.\ln\left[1-r_{\rm TE}^{(1)}(i\xi_l,k_{\bot})
r_{\rm TE}^{(2)}(i\xi_l,k_{\bot})\,e^{-2aq_l}\right]\right\}.
\label{eq1}
\end{eqnarray}
\noindent
Here, $k_B$ is the Boltzmann constant, $\xi_l=2\pi k_B Tl/\hbar$ with
$l=0,\,1,\,2,\,\ldots$ are the Matsubara frequencies, the prime
near the summation sign multiplies the term with $l=0$ by
a factor of 1/2,
$k_{\bot}$ is the modulus of the wave vector projection on the
plate (i.e., perpendicular to the   $z$-axis) and
\begin{equation}
q_l^2\equiv q_l^2(i\xi_l,k_{\bot})=k_{\bot}^2+
\frac{\xi_l^2}{c^2}.
\label{eq2}
\end{equation}
\noindent
The reflection coefficients for the two independent
polarizations of the electromagnetic field (transverse magnetic,
TM, and transverse electric, TE) are given by
\begin{eqnarray}
&&
r_{\rm TM}^{(n)}(i\xi_l,k_{\bot})=
\frac{\varepsilon_l^{(n)}q_l-k_l^{(n)}}{\varepsilon_l^{(n)}q_l+k_l^{(n)}},
\nonumber \\
&&
r_{\rm TE}^{(n)}(i\xi_l,k_{\bot})=
\frac{\mu_l^{(n)}q_l-k_l^{(n)}}{\mu_l^{(n)}q_l+k_l^{(n)}},
\label{eq3}
\end{eqnarray}
\noindent
where $\varepsilon_l^{(n)}\equiv\varepsilon^{(n)}(i\xi_l)$,
$\mu_l^{(n)}\equiv\mu^{(n)}(i\xi_l)$ with $n=1,\,2$ are the
dielectric permittivity and magnetic permeability of the first
and second plates, respectively,
calculated at the imaginary Matsubara frequencies, and
\begin{equation}
{k_l^{(n)}}^2\equiv {k^{(n)}}^2(i\xi_l,k_{\bot})=k_{\bot}^2+
\varepsilon_l^{(n)}\mu_l^{(n)}\frac{\xi_l^2}{c^2}.
\label{eq4}
\end{equation}
\noindent
Recently the Lifshitz formula (\ref{eq1}) for magnetodielectric
media was rigorously rederived \cite{63} under the same
assumptions, as formulated above, in the framework of
quantum field-theoretical scattering approach.

The Casimir force per unit area of the plates (i.e., the Casimir
pressure) is obtained from Eq.~(\ref{eq1}) by the negative differentiation
with respect to $a$,
\begin{eqnarray}
P(a,T)&=&-\frac{k_BT}{\pi}\sum\limits_{l=0}^{\infty}
{\vphantom{\sum}}^{\prime}\int_0^{\infty}q_lk_{\bot}dk_{\bot}
\left\{\left[\frac{e^{2aq_l}}{r_{\rm TM}^{(1)}(i\xi_l,k_{\bot})
r_{\rm TM}^{(2)}(i\xi_l,k_{\bot})}-1\right]^{-1}\right.
\nonumber \\
&&+\left.\left[\frac{e^{2aq_l}}{r_{\rm TE}^{(1)}(i\xi_l,k_{\bot})
r_{\rm TE}^{(2)}(i\xi_l,k_{\bot})}-1\right]^{-1}\right\}.
\label{eq5}
\end{eqnarray}

We  come now to the determination of the class of materials whose
magnetic properties may influence the Casimir force. It is common knowledge
that all materials possess  diamagnetic polarization, i.e., they are
magnetized  in direction opposite to the applied magnetic field.
For all substances the magnetic permeability is represented in the form
\begin{equation}
\mu(i\xi)=1+4\pi\chi(i\xi),
\label{eq5a}
\end{equation}
\noindent
where $\chi(i\xi)$ is the magnetic susceptibility calculated along
the imaginary frequency axis. The magnitude of $\chi(i\xi)$ is a
monotonously decreasing function of $\xi$. For diamagnets the
diamagnetic polarization determines their magnetic properties
so that \cite{32,33,34}
$\chi(0)<0$, $\mu(0)<1$ and $|\mu(0)-1|\sim 10^{-5}$.
{}From this it follows that magnetic properties of diamagnets cannot
influence the Casimir force and one can put $\mu_l=1$, $l=0,\,1,\,2,\,\ldots$
in computations using Eqs.~(\ref{eq1}) and (\ref{eq5}). Typical
diamagnets are such materials as Au, Si, Cu and Ag. It is important that
Au, Si and Cu were used in experiments on measuring the Casimir force
(see, e.g., papers \cite{12,30,31,38,39,40,41,42} and review of all
related experiments \cite{43}). Thus, it is justified to omit magnetic
properties of these materials when comparing the experimental data
with theory.

Materials possessing  paramagnetic polarization are magnetized in the
direction of an applied magnetic field. For  paramagnets in the broad
sense $\chi(0)>0$ and $\mu(0)>1$ and no additional conditions on
the character of the magnetic permeability apply \cite{34}.
Paramagnetic effects, if they are present, overpower the diamagnetic ones
and determine the type of the material. Paramagnets may consist of
microparticles which are paramagnetic magnetizable but have no
intrinsic magnetic moment (the Van Vleck polarization paramagnetism
\cite{44}). The respective $\chi(0)$ is, however, negligibly small.
Because of this the magnetic properties of Van Vleck paramagnets do
not influence the Casimir force.

Paramagnets may also consist of microparticles possessing an intrinsic
(permanent) magnetic moment (the orientational paramagnetism
\cite{32,33,34,44}). In the narrow sense,  magnetic materials with
$\chi(0)>0$ are called paramagnets if the interaction of magnetic
moments of their constituent particles is negligibly small. At sufficiently
high temperature all paramagnets are in fact paramagnets in the narrow
sense. Their magnetic susceptibility varies from about $10^{-7}$ to
about $10^{-4}$. When temperature decreases, there occurs a magnetic phase
transition \cite{45,46,47}. It happens at some critical temperature
$T_{cr}$ specific for each material (for different materials $T_{cr}$
may vary \cite{32,33,34,45,46,47} from a few K to more than thousand K).
However, for all paramagnets in the broad sense, with exception of only
ferromagnets, $\chi(0)$ remains as small as mentioned above and takes only
a bit larger values in the vicinity of  absolute zero temperature, $T=0\,$K.
This leads to the conclusion that magnetic properties of paramagnets
(with the single exception of ferromagnets) cannot markedly affect the
Casimir force acting between macroscopic bodies. Thus, when calculating
the Casimir free energy (\ref{eq1}) and pressure (\ref{eq5}) for
these materials, one can put $\mu_l=1$ in the reflection coefficients
(\ref{eq3}) for all $l\geq 0$.

The subdivision of paramagnetic materials called {\it ferromagnets}
requires special attention with respect to the Casimir force.
For such materials $\mu(0)\gg 1$ at $T<T_{cr}$. In this case the latter
is referred to as the {\it Curie} temperature, $T_{cr}\equiv T_C$.
There is a lot of ferromagnetic materials with various electric
properties (both metals and dielectrics) \cite{48}. They are
characterized by strong interaction between constituent microscopic
magnetic moments which results in large values of $\mu$ at low
frequencies and in the possibility of  spontaneous magnetization
({\it hard} ferromagnetic materials). It is not reasonable to consider
parallel plates made of hard ferromagnetic materials  because the
magnetic interaction between such plates far exceeds any conceivable
Casimir force. Below we consider the so-called {\it soft} ferromagnetic
materials   which do not possess  spontaneous magnetization.
It is well known that the magnetic permeability of ferromagnets depends
on the applied magnetic field \cite{32,33,34}.
As a result the relation between {\boldmath$B$} and {\boldmath$H$}
used in the derivation of Eq.~(\ref{eq1}) becomes nonlinear and
depends on the history of the material (the so-called
{\it hysteresis}).
In the Casimir related
problems, however, no external magnetic field is applied to
material plates whereas the mean value of the fluctuating magnetic
field is equal to zero.
Because of this, here we consider what is often
referred to as {\it initial} permeability, i.e.,
$\mu(\mbox{\boldmath$H$}=0)$.
Thus one can continue in using linear relation
between {\boldmath$B$} and {\boldmath$H$} and apply
Eq.~(\ref{eq1}) to soft ferromagnetic materials (i.e. to materials
with $\mu\gg 1$) as was done, for instance, in
papers \cite{23,27,28,29,63}.
It is pertinent to note that more theoretical work should be
done in order to finally justify the applicability of the
Lifshitz formula to ferromagnetic materials with
$\mu\gg 1$, especially to hard ferromagnets.

An important question arising in the calculation of the Casimir force
between ferromagnetic plates is how quickly the initial magnetic
permeability decreases with the increase of frequency. The rate of
decrease of $\mu(i\xi)$ with increasing $\xi$ depends on the value
of electric resistance. The lower is the resistance of a ferromagnetic
material, the lower is the frequency at which
$\mu(i\xi)$ drops toward unity \cite{32,33,34}.
Thus, for ferromagnetic metals $\mu(i\xi)$ becomes equal to unity at
frequencies above of order $10^5\,$Hz (see, e.g., \cite{49})
and for ferromagnetic dielectrics at frequencies above of order $10^9\,$Hz
(see, e.g., \cite{50}). The first Matsubara frequency $\xi_1$
at $T=300\,$K is of order $10^{14}\,$Hz. Thus, $\xi_1$ is much larger
than the frequencies where magnetic permeability of ferromagnets drops
to unity. Because of this, in all applications of the Lifshitz formulas
(\ref{eq1}) and (\ref{eq5}) at room temperature (and even at much
lower temperatures) one can put $\mu_l=1$ at all $l\geq 1$ and include
ferromagnetic properties only in the zero-frequency term with $l=0$.
Keeping in mind that the contribution of the zero-frequency term
(and thereby magnetic properties) increases with the increase of
separation between the plates, below we perform all computations in
the region from $0.5\,\mu$m to $6\,\mu$m. Near the left boundary of this
interval the contribution of the zero-frequency term
 is of order of a few percent and at the right boundary this
term determines the total values of the Casimir
free energy and pressure (at larger separations the Casimir interaction
becomes too small to be measured).

In addition to the magnetic permeability, one needs
to know the frequency-dependent
dielectric permittivity for the materials under consideration in order to
compute the Casimir free energy and pressure. For metals, at separations above
$0.5\,\mu$m the contribution of the interband transitions into the Casimir
interaction is negligible. At such separations interaction is completely
determined by the role of free conduction electrons. Main approaches
to the calculation of the Casimir force between metal plates used
in the literature describe conduction electrons by means of the Drude
model \cite{6,14,35,51,52} or the plasma model \cite{14,36,37,43,53}.
Within the Drude model approach the dielectric permittivity along the
imaginary frequency axis is given by
\begin{equation}
\varepsilon_D(i\xi)=1+\frac{\omega_p^2}{\xi(\xi+\gamma)},
\label{eq6}
\end{equation}
\noindent
where $\omega_p$ is the plasma frequency and $\gamma$ is the
relaxation parameter. The dielectric permittivity of the plasma model
is obtained from (\ref{eq6}) by putting $\gamma=0$.
\begin{equation}
\varepsilon_p(i\xi)=1+\frac{\omega_p^2}{\xi^2}.
\label{eq7}
\end{equation}
Both models lead to markedly different theoretical predictions
for the Casimir pressure between two metal plates. Predictions based
on the Drude model have been experimentally excluded at high confidence
level in  experiments using a micromechanical torsional oscillator
\cite{14,30,31,42,43,54}. Below, however, we consider both permittivities
(\ref{eq6}) and (\ref{eq7}) on equal terms in order to obtain respective
 consequences on the role of magnetic properties in the framework of the
proposed models. We will suggest new experimental tests which could shed
additional light on the applicability of these models in the theory of
thermal Casimir force.

The permittivity of dielectric materials along the imaginary frequency
axis is described in the framework of the oscillator model \cite{55},
\begin{equation}
\varepsilon_d(i\xi)=1+\sum\limits_{j=1}^{K}
\frac{g_j}{\omega_j^2+\xi^2+\gamma_j\xi},
\label{eq8}
\end{equation}
where $\omega_j\neq 0$ are the oscillator frequencies, $g_j$ are the
oscillator strengths, $\gamma_j$ are the relaxation parameters,
and $K$ is the number of oscillators.

\section{Distance dependence of the Casimir force for
ferromagnetic metals}

Here, we consider the Casimir interaction between two similar parallel
plates made of ferromagnetic metal. We perform computations
in order to investigate the role of magnetic properties for both the
Casimir free energy per unit area and pressure. Keeping in mind the
proximity force approximation \cite{14}, this allows one to apply
the obtained results to the experimental configurations of a sphere
above a plate and of two parallel plates, respectively.

Let us consider the reflection coefficients (\ref{eq3}) for two plates
made of ferromagnetic metal at room temperature $T=300\,$K.
In accordance with Sec.\ II, magnetic properties
may contribute only at zero frequency. For the TM polarization of the
electromagnetic field we arrive at
\begin{equation}
r_{\rm TM}^{(n)}(0,k_{\bot})=1,
\label{eq9}
\end{equation}
\noindent
where $n=1,\,2$, i.e., the same result as for ordinary (nonmagnetic) metals.
For the TE polarization we arrive at different expressions depending on the
model of dielectric permittivity used. Thus, for the Drude model (\ref{eq6})
from Eq.~(\ref{eq3}) one obtains:
\begin{equation}
r_{{\rm TE},D}^{(n)}(0,k_{\bot})\equiv r_{{\rm TE},D}^{(n)}(0)=
\frac{\mu(0)-1}{\mu(0)+1}.
\label{eq10}
\end{equation}
\noindent
Alternatively, for the plasma model (\ref{eq7})
Eq.~(\ref{eq3}) leads to
\begin{equation}
r_{{\rm TE},p}^{(n)}(0,k_{\bot})=
\frac{\mu(0)ck_{\bot}-
\sqrt{c^2k_{\bot}^2+\mu(0)\omega_p^2}}{\mu(0)ck_{\bot}+
\sqrt{c^2k_{\bot}^2+\mu(0)\omega_p^2}}.
\label{eq11}
\end{equation}
\noindent
The magnitude of this reflection coefficient depends on the relationship
between $\mu(0)$ and $\omega_p$.

First we present the computational results for the Casimir interaction
of two plates made of the ferromagnetic metal Co with \cite{48}
$\mu_{\rm Co}(0)=70$.
Computations were performed at room
temperature using Eq.~(\ref{eq1}) (the
Casimir free energy per unit area of the plates) and Eq.~(\ref{eq5}) (the
Casimir pressure). In all terms of these equations with $l\geq 1$ we
put $\mu(i\xi_l)=1$ in accordance with the results of Sec.\ II.
In the zero-frequency terms, Eqs.~(\ref{eq9}) and (\ref{eq10}) or
(\ref{eq11}) have been used depending on the chosen model of
$\varepsilon$ (Drude or plasma). For Co one has \cite{56}
$\omega_{p,{\rm Co}}=3.97\,$eV and $\gamma_{\rm Co}=0.036\,$eV.
Below the computational results are presented as ratios to the
zero-temperature Casimir
energy per unit area and the Casimir pressure between two nonmagnetic
parallel plates made of ideal metal,
\begin{equation}
E_0(a)=-\frac{\pi^2}{720}\,\frac{\hbar c}{a^3}, \qquad
P_0(a)=-\frac{\pi^2}{240}\,\frac{\hbar c}{a^4}.
\label{eq12}
\end{equation}

In Fig.~1, the solid lines show the values of ${\cal F}_{\rm Co}/{E_0}$ (a)
and of $P_{\rm Co}/P_0$ (b) as functions of separation computed using the
dielectric permittivity of the Drude model (\ref{eq6}). In the same figure,
the dashed lines show the computational results obtained
with omitted magnetic properties of
Co [i.e., with $\mu(0)=1$]. Quantitatively, the role
of magnetic properties can be characterized by the ratio
\begin{equation}
\eta_{{\cal F},{\rm Co}}=\frac{{\cal F}_{\rm Co}^{\rm solid}-
{\cal F}_{\rm Co}^{\rm dashed}}{{\cal F}_{\rm Co}^{\rm dashed}}
\label{eq13}
\end{equation}
\noindent
and by a similarly defined quantity $\eta_{P,{\rm Co}}$. With the increase
of separation distance from $a=0.5\,\mu$m to $2\,\mu$m and then to
$6\,\mu$m, $\eta_{{\cal F},{\rm Co}}^{D}$ varies from 17\% to 63\% and to 93\%,
respectively. At the same separations $\eta_{P,{\rm Co}}^{D}$
takes the following
respective values: 12\%, 44\% and 92\%. This permits us to conclude
that when the Drude model is used to describe the dielectric properties
of a ferromagnetic metal, the magnetic properties markedly (up to two
times at large separations) increase the magnitude of the Casimir free
energy and pressure.

In Fig.~2(a,b) similar results for two Co plates described by the plasma
model (\ref{eq7}) are presented. The same notation as in Fig.~1 is used.
As is seen in Fig.\ 2, for ferromagnetic metal described by the plasma
model the impact of magnetic properties on the Casimir interaction is not
so pronounced, as in Fig.\ 1. Quantitatively, from Fig.\ 2(a) it follows
that at separations 0.5, 2 and $6\,\mu$m  $\eta_{{\cal F},{\rm Co}}^{p}$
varies from --8.9\% to --17\% and --10\%, respectively. {}From Fig.\ 2(b)
one finds that the values of $\eta_{P,{\rm Co}}^{p}$ at the same separations
are --6\%, --17\% and --14\%. Thus, if the plasma model is used, the
inclusion of magnetic properties decreases the magnitudes of the Casimir
free energy and pressure.  It is important to note that the dashed lines in
Fig.\ 2(a,b) are very close to the solid lines in Fig.\ 1(a,b) (the
relative differences are below 2.5\%). This means that experimentally it is
hard to resolve between the case when the metal of the plates is described by
the Drude model and magnetic properties influence the Casimir interaction
and the case when metal is described by the plasma model but magnetic
properties have no impact on the Casimir interaction.

Another ferromagnetic metal is Fe. We consider the role of magnetic
interactions for two parallel plates made of Fe with the
parameters \cite{48,56}
$\mu_{\rm Fe}(0)=10^4$,
$\omega_{p,{\rm Fe}}=4.09\,$eV and $\gamma_{\rm Fe}=0.018\,$eV.
Numerical computations were performed as described above using
Eqs.~(\ref{eq1}) and (\ref{eq5}). The computational results for
the Casimir free energy (a) and pressure (b) obtained on the basis of
the Drude model approach at $T=300\,$K
are presented in Fig.\ 3. As above, the solid lines
are computed taking into account the magnetic properties of Fe and
dashed lines with magnetic properties disregarded. As is seen in Fig.\ 3,
magnetic properties significantly increase the magnitudes of the Casimir
free energy and pressure.  Thus, at $a=0.5$, 2 and $6\,\mu$m the respective
correction factors vary as $\eta_{{\cal F},{\rm Fe}}^{D}=18$\%, 68\%, 100\%
and $\eta_{P,{\rm Fe}}^{D}=13$\%, 47\%, 99\%. In Fig.\ 4(a,b) similar
computational results for the two Fe plates are presented when the plasma
model is used for the description of dielectric properties. It can be seen
that for Fe described by the plasma model the influence of magnetic
properties on the Casimir interaction is much stronger than for Co using
the same model. For separations $a=0.5$, 2 and $6\,\mu$m respective
values of the correction factors are:
$\eta_{{\cal F},{\rm Fe}}^{p}=-3.5$\%, --31\%, --42\%
and $\eta_{P,{\rm Fe}}^{p}=0.21$\%, --21\%, --45\%.

In the limiting case of large separations the Casimir interaction between
two plates made of ferromagnetic metal can be found analytically.
In this case the zero-frequency term alone determines the total result.
When dielectric properties are described by the Drude model, both
reflection coefficients at zero frequency (\ref{eq9}) and (\ref{eq10})
do not depend on $k_{\bot}$. Substituting (\ref{eq9}) and (\ref{eq10})
into Eqs.\ (\ref{eq1}) and (\ref{eq5}) and preserving only the terms
with $l=0$, one arrives at
\begin{eqnarray}
&&
{\cal F}_{D}(a,T)=-\frac{k_BT}{16\pi a^2}\left\{
\zeta(3)+\mbox{Li}_3\left[\left(\frac{\mu(0)-1}{\mu(0)+1}\right)^2\right]
\right\},
\nonumber \\
&&
P_{D}(a,T)=-\frac{k_BT}{8\pi a^3}\left\{
\zeta(3)+\mbox{Li}_3\left[\left(\frac{\mu(0)-1}{\mu(0)+1}\right)^2\right]
\right\},
\label{eq14}
\end{eqnarray}
\noindent
where $\zeta(z)$ is the Riemann zeta function and $\mbox{Li}_n(z)$ is
the polylogarithm function. Note that at $a=6\,\mu$m Eq.~(\ref{eq14})
leads to the same values of the Casimir free energy and pressure as those
computed in Figs.\ 1 and 3. Using the equalities
\begin{equation}
\lim\limits_{z\to 1}\,\mbox{Li}_3(z)=\zeta(3), \qquad
\lim\limits_{z\to 0}\,\mbox{Li}_3(z)=0,
\label{eq15}
\end{equation}
\noindent
we easily obtain from (\ref{eq14}) the asymptotic results for the
case of very high magnetic permeability $\mu(0)\gg 1$,
\begin{equation}
{\cal F}_{D}(a,T)=-\frac{k_BT}{8\pi a^2}\zeta(3),
\qquad
P_{D}(a,T)=-\frac{k_BT}{4\pi a^3}\zeta(3),
\label{eq16}
\end{equation}
\noindent
and for nonmagnetic Drude metals
\begin{equation}
{\cal F}_{D}(a,T)=-\frac{k_BT}{16\pi a^2}\zeta(3),
\qquad
P_{D}(a,T)=-\frac{k_BT}{8\pi a^3}\zeta(3).
\label{eq17}
\end{equation}
\noindent
The equalities in (\ref{eq16}) coincide with respective results
obtained for the nonmagnetic metals described by the plasma model
in the limit of large separations (the standard ideal-metal results
\cite{14}). Similar approximate equalities were noted above on the
basis of computations performed at shorter separations.
As to Eq.~(\ref{eq17}), it coincides with the prediction of the Drude
model approach for nonmagnetic metals at large separations
\cite{6,14,35,51,52}, as it should be.

For plates made of ferromagnetic metal described by the plasma model
the asymptotic behavior of the Casimir interaction at large separations
is a bit more cumbersome. For brevity, we restrict ourselves by the
calculation of the Casimir free energy. Introducing the dimensionless
variable $y=2ak_{\bot}$, we can rearrange Eq.~(\ref{eq11}) to the form
\begin{equation}
r_{{\rm TE},p}^{(n)}(0,y)=
\frac{\mu(0)y-\sqrt{y^2+\mu(0)\tilde{\omega}_p^2}}{\mu(0)y+
\sqrt{y^2+\mu(0)\tilde{\omega}_p^2}},
\label{eq18}
\end{equation}
\noindent
where the dimensionless plasma frequency is defined as
\begin{equation}
\tilde{\omega}_p=\frac{\omega_p}{\omega_c}\equiv\frac{2a\omega_p}{c}.
\label{eq19}
\end{equation}
\noindent
The reflection coefficient (\ref{eq18}) can be expanded in powers of
small parameter,
\begin{equation}
\alpha\equiv\frac{1}{\tilde{\omega}_p}=
\frac{\lambda_p}{2\pi}\,\frac{1}{2a}\equiv\frac{\delta_0}{2a}\ll 1,
\label{eq20}
\end{equation}
\noindent
where $\delta_0$ is the penetration depth of the electromagnetic
oscillations into a metal described by the plasma model.
Using (\ref{eq20}) the reflection coefficient (\ref{eq18}) can be
represented as
\begin{equation}
r_{{\rm TE},p}^{(n)}(0,y)\approx
\frac{\sqrt{\mu(0)}\alpha y-1}{\sqrt{\mu(0)}\alpha y+1}.
\label{eq21}
\end{equation}
\noindent
Substituting this into the term of Eq.~(\ref{eq1}) with $l=0$
rearranged using the variable $y$, one obtains
\begin{equation}
{\cal F}_p(a,T)=\frac{k_BT}{16\pi a^2}\left\{-\zeta(3)+
\int_0^{\infty}ydy\ln\left[1-\left(
\frac{\sqrt{\mu(0)}\alpha y-1}{\sqrt{\mu(0)}\alpha y+1}\right)^2\,
e^{-y}\right]\right\}.
\label{eq22}
\end{equation}

Further simplification of Eq.~(\ref{eq22}) is possible under a condition
$\sqrt{\mu(0)}\alpha\ll 1$ readily satisfied at separations above
$6\,\mu$m for nearly all magnetic materials. Expanding under the integral
in powers of the small parameter $\sqrt{\mu(0)}\alpha$ and preserving only
the first-order contribution, we arrive at
\begin{equation}
{\cal F}_p(a,T)=\frac{k_BT}{8\pi a^2}\left[-\zeta(3)+
2\sqrt{\mu(0)}\alpha\int_0^{\infty}dy\frac{y^2}{e^{y}-1}\right].
\label{eq23}
\end{equation}
\noindent
After the integration in Eq.~(\ref{eq23}) is done, the result is
\begin{equation}
{\cal F}_p(a,T)=-\frac{k_BT}{8\pi a^2}\zeta(3)\left[1
-2\sqrt{\mu(0)}\frac{\delta_0}{a}\right].
\label{eq24}
\end{equation}
\noindent
For nonmagnetic metals $\mu(0)=1$ and Eq.~(\ref{eq24}) coincides with
the previously obtained result for the high-temperature Casimir free
energy in the case of metals described by the plasma model \cite{14,37}.
As can be seen from Eq.~(\ref{eq24}), the account of magnetic properties of
a ferromagnetic metal described by the plasma model decreases the
magnitude of the Casimir free energy, as was already shown above by means of
numerical computations. The values of ${\cal F}_p$ at $a=6\,\mu$m
calculated using Eq.~(\ref{eq24}) coincide with those computed in
Figs.\ 2(a) and 4(a).

\section{Distance dependence of the Casimir force for a ferromagnetic
metal interacting with a nonmagnetic metal}

In this section we consider the configuration of two dissimilar
parallel plates one of which is made of ferromagnetic metal and the other of
nonmagnetic metal. Note that to determine the role of magnetic
properties it would be not informative to consider the interaction of the
first plate made of ferromagnetic metal with the second plate made of an
ordinary (nonmagnetic) dielectric. The point is that, in accordance with
Eqs.\ (\ref{eq9})--(\ref{eq11}), magnetic properties contribute to the
Casimir interaction only through the transverse electric mode at $\xi=0$.
However, the substitution of Eq.~(\ref{eq8}) into Eq.~(\ref{eq3}) leads to
\begin{equation}
r_{{\rm TE},d}^{(2)}(0,k_{\bot})=0.
\label{eq25}
\end{equation}
\noindent
As a result, the magnetic properties of a ferromagnetic metal plate
interacting with a plate made of nonmagnetic dielectric do not contribute
into the Lifshitz formula.

The reflection coefficients for the plate  made of ferromagnetic metal
(Co or Fe) at zero frequency are given by Eqs.\ (\ref{eq9})--(\ref{eq11})
with $n=1$ depending on the  model of dielectric permittivity used.
At nonzero Matsubara frequencies the reflection coefficients for this plate
are given by Eq.~(\ref{eq3}) with $n=1$ and $\mu_l^{(1)}=1$
($l=1,\,2,\,\ldots$). Equation (\ref{eq3}) with $n=2$ and $\mu_l^{(2)}=1$
for all $l=0,\,1,\,2,\,\ldots$ also determines the reflection coefficients
for the plate made of an ordinary (nonmagnetic) metal. As a nonmagnetic
metal we use Au with the parameters \cite{56,57}
$\omega_{p,{\rm Au}}=9.0\,$eV,
$\gamma_{\rm Au}=0.035\,$eV. All necessary parameters of Co
are listed in Sec.~III.

In Fig.~5 we present the computational results (the solid lines) for the
Casimir free energy (a) and pressure (b) as functions of separation
computed for the configuration
of Co-Au plates by Eqs.~(\ref{eq1}) and (\ref{eq5})
using the Drude model approach. The same notation as in Figs.\ 1--4 is used.
However, in this case the dashed lines, computed with the magnetic
properties of Co disregarded, coincide with the solid lines. The reason is
that for Au described within the Drude model it holds
\begin{equation}
r_{{\rm TE},D}^{(2)}(0,k_{\bot})=0
\label{eq26}
\end{equation}
\noindent
[compare with Eq.~(\ref{eq10}) where $\mu(0)=1$].
As a result, similar to the case when the second plate is made of
nonmagnetic dielectric, the magnetic properties of Co do not contribute
to the Casimir free energy and pressure.

Another situation holds when metals are described by means of the
plasma model (\ref{eq7}). The computational results
at $T=300\,$K are shown in Fig.\ 6
for the Casimir free energy (a) and pressure (b). In the same way,
as in Figs.\ 2 and 4,  the dashed lines computed with the magnetic
properties disregarded lie above the solid lines. However, quantitatively
the role of magnetic properties is rather moderate. Thus, at separations
$a=0.5$, 2 and $6\,\mu$m
$\eta_{{\cal F},{\rm Co-Au}}^{p}=-8.2$\%, --11\%, and --5.5\%, respectively,
whereas $\eta_{P,{\rm Co-Au}}^{p}=-6.9$\%, --12\%, and --7.9\%, respectively.
These relative differences are in fact rather close to the
relative differences between the Casimir free energy per unit
area and pressure
in the configuration of Co-Au plates computed using the Drude and the
plasma model approaches with magnetic properties of Co disregarded
(--12\% for the free energy and --8.4\% for the pressure at the
shortest separation $a=0.5\,\mu$m).

As one more configuration we consider the plate made of
ferromagnetic Fe interacting with the Au plate. When the Drude model
is used the computational results are presented in Fig.\ 7(a,b) with
the same notation as above (the parameters of Fe are listed in Sec.\ III).
Here, the magnetic properties of Fe do not influence  the results
obtained. When the plasma model is used in the computations, the impact of
the magnetic properties of Fe on the obtained results is rather
pronounced. In Fig.\ 8 the Casimir free energy (a) and pressure (b)
at $T=300\,$K
are shown as functions of separation. Here, the solid lines taking
magnetic properties into account deviate significantly from the dashed
lines computed with magnetic properties of Fe disregarded. At separations
$a=0.5$, 2 and $6\,\mu$m the above quantitative characteristics of the role
of magnetic properties take the following values:
$\eta_{{\cal F},{\rm Fe-Au}}^{p}=-19$\%, --46\% and --38\%, respectively, and
$\eta_{P,{\rm Fe-Au}}^{p}=-13$\%, --41\% and --48\%, respectively.
This is larger in magnitude (or nearly equal for $\eta_{P,{\rm Fe-Au}}^{p}$
at $a=6\,\mu$m) than the relative differences in the Casimir free
energy and pressure in the configuration of Fe-Au plates computed using
the Drude and the plasma model approaches with magnetic properties of
Fe disregarded. The relevance of the configuration of
a ferromagnetic metal plate interacting with a nonmagnetic metal
plate for future
experiments is discussed below (see Sec.\ VII).

At large separation distances ($a\geq 6\,\mu$m) the analytical
representations for the Casimir free energy in the configuration of
a ferromagnetic metal plate near a nonmagnetic
metal plate can be obtained.
When the Drude model is used, the result is given by the TM contribution
to the zero-frequency term of Eq.~(\ref{eq1}) presented  in Eq.~(\ref{eq17}).
This is because the TE contribution vanishes due to Eq.~(\ref{eq26})
valid for the plate made of a nonmagnetic Drude metal. When the plasma
model is used, we can use  expression (\ref{eq21}) with $n=1$ and
$\alpha=\alpha_1\equiv 1/\tilde{\omega}_{p,{\rm Fe}}=c/(2a\omega_{p,{\rm Fe}})$
for the TE reflection coefficient of the plate made of ferromagnetic metal.
Under the same condition (\ref{eq20}) for a nonmagnetic metal ($n=2$),
Eq.\ (\ref{eq21}) results in
\begin{equation}
r_{{\rm TE},p}^{(2)}(0,y)\approx -1+2\alpha_2y,
\label{eq27}
\end{equation}
\noindent
where
$\alpha=\alpha_2\equiv 1/\tilde{\omega}_{p,{\rm Au}}=
c/(2a\omega_{p,{\rm Au}})$.

Substituting Eq.\ (\ref{eq21}) with $n=1$ and Eq.\ (\ref{eq27}) into
the zero-frequency term of Eq.\ (\ref{eq1}) with account of Eq.\ (\ref{eq26}),
we obtain
\begin{equation}
{\cal F}_{p}(a,T)=\frac{k_BT}{16\pi a^2}\left\{-\zeta(3)+
\int_0^{\infty}ydy\ln\left[1+\left(
\frac{\sqrt{\mu(0)}\alpha_1y-1}{\sqrt{\mu(0)}\alpha_1y+1}\right)
(1-2\alpha_2y)e^{-y}\right]\right\}.
\label{eq28}
\end{equation}
\noindent
Using also the condition $\sqrt{\mu(0)}\alpha_1\ll 1$ and restricting
ourselves by the first-order perturbation theory in the small parameters
$\sqrt{\mu(0)}\alpha_1$ and $\alpha_2$, we arrive at the result
\begin{equation}
{\cal F}_{p}(a,T)=-\frac{k_BT}{8\pi a^2}\zeta(3)\left[1-
\frac{\sqrt{\mu(0)}\delta_{01}+\delta_{02}}{a}\right],
\label{eq29}
\end{equation}
\noindent
where $\delta_{01}$ and $\delta_{02}$ are the relative penetration depths
of the electromagnetic oscillations in the first and second plates,
respectively, defined in accordance with Eq.\ (\ref{eq20}).
{}From Eq.\ (\ref{eq29}) it is seen that the account of magnetic properties
of the ferromagnetic metal in the framework of the plasma model makes the
magnitude of the Casimir free energy smaller. This is in accordance with
the computational results in Figs.\ 6(a) and 8(a). The values of the
Casimir free energy at $a=6\,\mu$m calculated from Eq.\ (\ref{eq29})
fit the respective computational results in Figs.\ 6(a) and 8(a).

\section{Distance dependence of the Casimir force for ferromagnetic
dielectrics}

Ferromagnetic dielectrics are very prospective for the investigation
of the impact of magnetic properties on the Casimir force. There are
many materials which, while displaying physical properties characteristic
for dielectrics, demonstrate ferromagnetic behavior under the influence
of an external magnetic field (see, e.g., review \cite{58}).
Many examples of such a substance are composite materials \cite{59,60}
 obtained on the
basis of a polymer compound with inclusion of nanoparticles of ferromagnetic
metals, different transition metal doped oxides \cite{60a}, etc.
In addition to numerous dielectric materials displaying
 ferromagnetic properties listed in \cite{58}, one could mention
 the Chromium Bromide \cite{60b}, films of ZnO doped \cite{60c} with
 magnetic ions of Mn and Co, and epitaxial CeO${}_2$ films doped
 by cobalt \cite{60d}.

Ferromagnetic dielectrics
are widely used in different magneto-optical devices. Numerical
computations of the Casimir interaction reported below
 are performed for the model of composite material on the basis
of polystyrene with the volume fraction of ferromagnetic metal particles
in the mixture $f=0.25$. The magnetic permeability of such kind of
materials may vary over a wide range \cite{59}. Below we use $\mu(0)=25$.
The dielectric permittivity of polystyrene $\varepsilon_d(i\xi)$ is
presented in the form (\ref{eq8}) with $K=4$ oscillators. The parameters
of oscillators $g_j$, $\omega_j$ and $\gamma_j$ are taken
from \cite{55,61}. Specifically, at zero frequency $\varepsilon_d(0)=2.56$.
The dielectric permittivity of the used ferromagnetic dielectric
is obtained as \cite{62}
\begin{equation}
\varepsilon_{fd}^{(n)}(i\xi)=\varepsilon_d(i\xi)\left(1+
\frac{3f}{1-f}\right),
\label{eq30}
\end{equation}
\noindent
which leads to $\varepsilon_{fd}^{(n)}(0)=5.12$.
The value of $f$ chosen above belongs to the range of validity of
this equation \cite{59}.

We start with the configuration of two similar plates ($n=1,\,2$) made of the
ferromagnetic dielectric with the parameters presented above.
As in previous sections, the computations are performed using
Eqs.\ (\ref{eq1}) and (\ref{eq5}) where the magnetic properties are
included in the zero-frequency term ($l=0$)
at $T=300\,$K. In all terms with $l\geq 1$ it
is assumed that $\mu_l=1$. At zero frequency the TM reflection
coefficient for a ferromagnetic dielectric plate is obtained from
Eq.\ (\ref{eq3}),
\begin{equation}
r_{\rm TM}^{(n)}(0,k_{\bot})\equiv r_{\rm TM}^{(n)}(0) =
\frac{\varepsilon^{(n)}(0)-1}{\varepsilon^{(n)}(0)+1}.
\label{eq31}
\end{equation}
\noindent
The TE reflection coefficient for a ferromagnetic dielectric plate
at $\xi=0$ coincides with that in Eq.\ (\ref{eq10}) for a ferromagnetic
metal described by the Drude model [compare with Eq.\ (\ref{eq25})
for a nonmagnetic dielectric].
The computational results for the Casimir free energy (a) and
pressure (b) as functions of separation are shown in Fig.\ 9.
The solid lines are computed with magnetic properties  taken into
account, the dashed lines are obtained with magnetic properties
disregarded ($\mu_l=1$ at all $l=0,\,1,\,2,\,\ldots$).
As can be seen in Fig.\ 9, the influence of magnetic properties on the
Casimir force increases with the increase of separation. Thus, at separations
$a=0.5$, 2 and $6\,\mu$m the parameter introduced in Eq.\ (\ref{eq13})
takes the values $\eta_{{\cal F},fd}=54$\%, 166\% and 203\%, respectively.
A similar situation holds for the Casimir pressure where
$\eta_{P,fd}=36$\%, 133\% and 203\% at the same respective separations.

As one more example, we consider the configuration of one plate made
of ferromagnetic dielectric ($n=1$) and the other plate made of a nonmagnetic
metal Au ($n=2$). Let the dielectric permittivity of Au,
$\varepsilon_{p,{\rm Au}}(i\xi)$, be described by the  plasma model (\ref{eq7})
with $\omega_p=9.0\,$eV. This choice is caused by the fact that when one
plate is made of a nonmagnetic Drude metal the magnetic properties of the
other plate do not influence  the Casimir interaction because of
Eq.\ (\ref{eq26}). In addition, the computational results for the Au plate
described by the Drude model interacting with the ferromagnetic
dielectric plate are nearly coinciding with those when Au is described by the
plasma model and the magnetic properties of ferromagnetic dielectric are
disregarded (see below).

The computational results are presented in Fig.\ 10 for the Casimir free
energy (a) and pressure (b). The solid
(dashed) lines show the results computed
using the plasma model for the Au plate with magnetic properties of the
ferromagnetic dielectric plate included
(disregarded). Note that if the Drude model
is used to describe the dielectric properties of the Au plate, the
obtained results nearly coincide with the dashed line within the range
of separations considered. The relative deviation between
the results obtained
using both models is equal to only 0.25\% and 0.09\% at separations
$a=0.5$ and $2\,\mu$m, respectively, and continues to decrease with the
increase of separation.

As can be seen in Fig.~10, there is the profound effect of  magnetic
properties of ferromagnetic dielectric
on the Casimir interaction in this configuration if Au is
described by the plasma model. Thus, at separations of 0.5, 2 and $6\,\mu$m
the respective values of $\eta_{{\cal F},fd-{\rm Au}}^{p}$ are equal to
--22\%, {--82\%}
and --111\%. For the Casimir pressure at the same
respective separations one has $\eta_{P,fd-{\rm Au}}^{p}=-14$\%, --60\% and
--110\%. What is more important, the Casimir free energy
${\cal F}$ changes sign and becomes positive (we remind that $E_0<0$) at
separations $a>2.9\,\mu$m [see Fig.\ 10(a)].
According to the proximity force approximation, the Casimir force
acting between a sphere of radius $R$ and a plate spaced at separation
$a\ll R$ from each other are approximately equal \cite{5,14} to
$2\pi R{\cal F}(a,T)$. This means that at separations
$a>2.9\,\mu$m the Casimir force acting between the sphere and the plate
is repulsive.

A similar situation takes place for the Casimir pressure. From Fig.\ 10(b)
it follows that at separations $a>3.8\,\mu$m the Casimir pressure $P$
changes its sign and becomes positive. This means that the Casimir force
acting between a ferromagnetic dielectric plate and Au plate described
by the plasma model becomes repulsive. We emphasize that the effect of
repulsion for the two parallel plates interacting through the vacuum gap
found by us is not analogous to the results \cite{28} discussed
in Introduction. The point is that the paper \cite{28} used some idealized
magnetodielectric materials of the plates with frequency-independent
$\varepsilon$ and $\mu$.
As was shown \cite{29}, the values of magnetic permeabilities
of real materials at characteristic frequencies contributing to the
Casimir force are much less than those required to obtain the effect of
repulsion because they quickly vanish with the increase of frequency.
In the asymptotic limit of very large separations, where the
zero-frequency $\varepsilon$ and $\mu$ can be used,
the repulsive Casimir force
was recently found \cite{63}  in the configuration
of an ideal metal cylinder above a magnetodielectric plate.
This result was obtained under the assumption that temperature is
equal to zero. The Casimir repulsion was predicted for the magnetic
permeability of the plate $\mu=100$ and dielectric permittivity
$\varepsilon<33$ or $\mu=10$ and $\varepsilon<4$. Thus in both
cases the materials of the plate are ferromagnetic dielectrics.
In contrast to this, we consider a real ferromagnetic dielectric plate
interacting with an Au plate at room temperature
and take into account the dependence
of their magnetic permeability and dielectric permittivity on the frequency.
Therefore the effect of repulsion found by us can be used as an
experimental test for the influence of magnetic properties on the
Casimir force and for the model of dielectric permittivity
of a metal plate (see Sec.\ VII).

Now we consider some analytical results that can be obtained in the
limiting case of large separations. For two similar plates made of
ferromagnetic dielectric one can use the reflection coefficient (\ref{eq10})
(as was noted above, this one is the same as for a ferromagnetic metal
described by the Drude model) and (\ref{eq31}). The resulting Casimir
free energy per unit area is given by
\begin{equation}
{\cal F}_{fd}(a,T)=-\frac{k_BT}{16\pi a^2}\left\{
\mbox{Li}_3\left[\left(
\frac{\varepsilon(0)-1}{\varepsilon(0)+1}\right)^2\right]
+\mbox{Li}_3\left[\left(\frac{\mu(0)-1}{\mu(0)+1}\right)^2\right]\right\},
\label{eq32}
\end{equation}
\noindent
where $\varepsilon(0)$, as defined in Eq.~(\ref{eq30}), and $\mu(0)$ are
the dielectric permittivity and magnetic permeability of the ferromagnetic
dielectric. If we have two dissimilar plates where one is made of
ferromagnetic dielectric and the other one of a nonmagnetic metal
described by the Drude model, the Casimir free energy per unit area is
determined by the contribution of the TM mode alone,
\begin{equation}
{\cal F}_{D}(a,T)=-\frac{k_BT}{16\pi a^2}
\mbox{Li}_3\left[\frac{\varepsilon(0)-1}{\varepsilon(0)+1}\right].
\label{eq33}
\end{equation}

For dissimilar plates where the metal plate is described by the plasma model,
with account of Eqs.\ (\ref{eq9}), (\ref{eq10}), (\ref{eq27}) and
(\ref{eq31}), one obtains
\begin{eqnarray}
&&
{\cal F}_{p}(a,T)=\frac{k_BT}{16\pi a^2}\left\{
-\mbox{Li}_3\left[\frac{\varepsilon(0)-1}{\varepsilon(0)+1}\right]
\right.
\label{eq34}\\
&&~~~~~~
\left.+
\int_{0}^{\infty}ydy\ln\left[1+\frac{\mu(0)-1}{\mu(0)+1}
(1-2\alpha_2y)e^{-y}\right]\right\}.
\nonumber
\end{eqnarray}
\noindent
By performing integration with respect to $y$ we arrive at
\begin{eqnarray}
&&
{\cal F}_{p}(a,T)=-\frac{k_BT}{16\pi a^2}\left\{
\mbox{Li}_3\left[\frac{\varepsilon(0)-1}{\varepsilon(0)+1}\right]
\right.
\label{eq35}\\
&&~~~~~~
\left.+
\mbox{Li}_3\left[\frac{1-\mu(0)}{1+\mu(0)}\right]
\left(1-2\frac{\delta_{02}}{a}\right)
\right\},
\nonumber
\end{eqnarray}
\noindent
where the penetration depth of the electromagnetic oscillations into Au,
$\delta_{02}$, is defined in accordance with Eq.\ (\ref{eq20}).

The Casimir free energy in Eq.~(\ref{eq35}) can be both negative and positive
leading to the attractive and repulsive Casimir force, respectively,
in the configuration of a sphere above a plate used in most of recent
experiments \cite{14,43}. Keeping in mind that $\delta_{02}\ll a$,
the Casimir free energy is negative if the following condition is
satisfied:
\begin{equation}
\mbox{Li}_3\left[\frac{\varepsilon(0)-1}{\varepsilon(0)+1}\right]>
\left|\mbox{Li}_3\left[
\frac{1-\mu(0)}{1+\mu(0)}\right]\right|\,
\left(1-2\frac{\delta_{02}}{a}\right).
\label{eq36}
\end{equation}
\noindent
If, on the opposite, it holds
\begin{equation}
\mbox{Li}_3\left[\frac{\varepsilon(0)-1}{\varepsilon(0)+1}\right]<
\left|\mbox{Li}_3\left[\frac{1-\mu(0)}{1+\mu(0)}\right]
\right|\,\left(1-2\frac{\delta_{02}}{a}\right),
\label{eq37}
\end{equation}
\noindent
then the Casimir free energy is positive and the Casimir force acting
in the sphere-plate configuration is repulsive.

In Fig.~11 we show the region of attraction (below the solid line) and
repulsion (above the solid line) in the $[\varepsilon(0),\mu(0)]$-plane
at separation distance $a=6\,\mu$m. For points belonging to the
solid line the Casimir force acting between the sphere and the plate
vanishes [for the coordinates of these points the inequalities
(\ref{eq36}) and (\ref{eq37}) become equalities]. Keeping in mind
that for ferromagnetic dielectrics $\varepsilon(0)$ is typically not very
small (for the material discussed above it is equal to 5.12) the region of
the repulsive Casimir force is rather restricted. This is connected with the
fact that the solid line in Fig.\ 11 has the vertical asymptote
$\varepsilon(0)=8.45$. Thus, there is no repulsive Casimir force at
$a=6\,\mu$m in the sphere-plate configuration for ferromagnetic
dielectrics possessing larger values of $\varepsilon(0)$.
Note that although the analytic results (\ref{eq36}) and (\ref{eq37}) can
be used only at sufficiently large separations ($a\geq 6\,\mu$m),
the results of numerical computations presented in Fig.\ 10 show that
the Casimir repulsion due to magnetic properties of ferromagnetic dielectric
may exist at shorter separations as well.

\section{The Casimir force in the vicinity of Curie temperature}

As mentioned in Sec.~II, at the Curie temperature $T_C$ specific for each
material ferromagnets undergo a magnetic phase transition \cite{46,48}.
At higher temperatures they become paramagnets  in the narrow sense which
are characterized by negligibly small magnetic properties with respect
to the Casimir force. In this section we consider the behavior of the
Casimir free energy and pressure under the magnetic phase transition
which occurs with the increase of temperature in the configuration of
two similar plates made of ferromagnetic metals. As such a metal, here
we use Gd. The reason is that Co and Fe used in computations of
Secs.\ III and IV possess rather high Curie temperatures (1388\,K and 1043\,K,
respectively \cite{64}). Keeping in mind that it is hard to measure
the Casimir force at such high temperatures, we consider Gd which Curie
temperature  is of about 290\,K depending on the treatment of a sample
(see, e.g., \cite{65,66}). In the literature, Gd is often discussed
in connection with its ferromagnetic properties, and the admixtures
of Gd atoms are included in different materials (see, e.g., \cite{67,68}).
The Drude parameters of Gd are equal \cite{69}
to $\omega_{p,{\rm Gd}}=9.1\,$eV,
$\gamma_{\rm Gd}=0.58\,$eV.

Computations of the Casimir free energy and pressure in the configuration
of two Gd plates as functions of temperature in the vicinity of Curie
temperature require  respective values of $\mu(0)$ for Gd
at $T<T_C$ [at $T>T_C$, $\mu_{\rm Gd}(0)=1$ to  high accuracy]. In Fig.\ 12,
using the data \cite{68}, we model the approximate dependence
of $\mu_{\rm Gd}(0)$ in the temperature region from 280\,K to 300\,K.
Then the Casimir free energy and pressure were computed as functions
of temperature using Eqs.\ (\ref{eq1}) and (\ref{eq5}) with above values of
the Drude parameters. The computational results for the Casimir free
energy are presented in Fig.\ 13(a) and for the Casimir pressure in
Fig.\ 13(b) at separation $a=500\,$nm. In both figures (a) and (b)
the solid and dashed lines marked 1 and 2 indicate the results computed
using the Drude and plasma model for the characterization of the dielectric
permittivity of Gd, respectively. As in previous sections, the solid
lines take into account the magnetic properties of Gd. The dashed lines
were computed with magnetic properties disregarded. As can be seen in
Fig.\ 13(a,b), at $T>T_C$ the magnetic properties do not influence the
Casimir free energy and pressure. At the same time, the Drude and plasma
model approaches lead to results differing for about --23.4\% for the Casimir
free energy and --19.5\% for the Casimir pressure.

The computational results at $T<T_C$ are of special interest. Here, the
magnetic properties influence  the Casimir free energy and pressure.
Below of about 288\,K this influence is almost temperature-independent.
Quantitatively, at $T=280\,$K the relative influence of magnetic
properties on the Casimir free energy is
$\eta_{{\cal F},{\rm Gd}}^{D}=11.6$\% if the Drude model is used and
$\eta_{{\cal F},{\rm Gd}}^{p}=-3.6$\% if computations are done by means of
the plasma model. Similar  situation holds for the Casimir pressure.
Here, the relative influence of magnetic properties is characterized by
$\eta_{P,{\rm Gd}}^{D}=7.4$\% for the Drude model and
$\eta_{P,{\rm Gd}}^{p}=-3.3$\% for the plasma model.
With account of magnetic properties, the relative difference between
the predictions of the Drude and plasma model approaches at $T=280\,$K is
approximately equal to --6.2\% for the Casimir free energy and --7\%
for the Casimir pressure. Thus, the magnetic phase transition provides
additional opportunities for the investigation of the impact of magnetic
properties on the Casimir force and for the selection between different
theoretical approaches to the thermal Casimir force.

\section{Conclusions and discussion}

In the foregoing we have investigated the possible impact of magnetic
properties of real materials on the thermal Casimir force in the
configuration of two parallel plates. This was done in the framework of
the Lifshitz theory of dispersion forces generalized for magnetodielectric
media described by the frequency-dependent dielectric permittivity
and magnetic permeability. The dielectric permittivity of metals was
described in the framework of both the Drude and the plasma model
approaches suggested in the literature for the calculation of the Casimir
force at nonzero temperature.

It was concluded that magnetic properties of all diamagnetic materials and
of paramagnetic materials in the broad sense with the single exception of
ferromagnets do not influence on  Casimir force.
As to ferromagnets, the influence of their magnetic properties on
the Casimir force is performed solely through the contribution of the
zero-frequency term in the Lifshitz formula.
Detailed calculations
of the thermal Casimir force have been performed for the following
configurations: two ferromagnetic metal plates; one plate made of
ferromagnetic metal and the other plate made of nonmagnetic metal;
two plates made of ferromagnetic dielectric;  one plate made of
ferromagnetic dielectric and the other plate made of nonmagnetic metal.
In some cases the relative
differences due to account of magnetic properties
were shown to achieve several tens and even hundreds
of percent. It was shown also that the
impact of magnetic properties on the
Casimir force may be quite different (or even absent) depending on
whether the Drude or the plasma model description of the dielectric
permittivity of metals is used.

The possible influence of magnetic properties of ferromagnets on the
Casimir force may be considered somewhat analogous to the proposed
influence of real drift current of conduction electrons. If it is
assumed that the fluctuating electromagnetic field can initiate such
a current, we arrive to the Drude model approach to the thermal
Casimir force which is considered as the most natural one by some of the
authors \cite{35,51,52}. This approach, however,
was found to be in drastic contradiction with the results of several
precision experiments \cite{14,30,31,42,54}. Because of this the problem
arises whether the fluctuating electromagnetic field can lead to
magnetic effects in ferromagnets. This problem awaits for its
experimental resolution.

The possibility to obtain the effect of the Casimir repulsion between
two magnetodielectric plates separated with a vacuum gap was analyzed
taking into account real material properties. It was shown that the
model of magnetic materials with frequency-independent $\varepsilon$
and $\mu$ used in the literature to obtain such a repulsion is
inadequate.  For real materials with frequency-dependent $\varepsilon$
and $\mu$ it is not possible to obtain the Casimir repulsion in the
configuration of two plates made of ferromagnetic dielectrics or
ferromagnetic metals described by the Drude model. According to our
results, a configuration demonstrating the Casimir repulsion due to magnetic
properties is the dissimilar pair of plates one of which is made of
ferromagnetic dielectric and the other one of nonmagnetic metal
described by the plasma model. This was shown both analytically and
numerically. It would be interesting to perform further numerical
studies of Casimir forces for different composite materials
in order to investigate in more detail the possibility of the
Casimir repulsion.

We now turn our attention to the discussion of feasible experiments which
could provide tests for the possible influence of magnetic properties of
ferromagnetic materials on the Casimir force and for the
used model of the
dielectric permittivity of metal (Drude or plasma). It would be most simple
to admit that the Drude model approach has already been excluded by
previous measurements \cite{30,31,42,54} and deal with only the magnetic
properties. Presently the most precise measurements of the Casimir pressure
at separations of $0.5\,\mu$m are performed by means of micromechanical
torsional oscillator \cite{30,31} (the experiments using an atomic force
microscope \cite{13,14} have the highest precision
at separations of about 100\,nm).
 Precise measurements of the Casimir interaction at
separations of a few $\mu$m are not yet available. {}From Fig.\ 2(b)
the relative difference between the solid and dashed lines at $a=0.5\mu$m
is equal to $\eta_{P,{\rm Co}}^{p}=-6.0$\%.
We keep in mind that in the experiment
\cite{31} the relative half-width of the confidence interval for the
difference between experimental and theoretical Casimir pressure
at $a=0.5\mu$m is equal to 2.8\% at a 95\% confidence level. Thus, the
experimental precision is sufficient to exclude one of the possibilities,
i.e., that the magnetic properties influence (or do not influence)  the
Casimir force.

It would be more interesting, however, to experimentally verify both
options (i.e., that the magnetic properties influence or
do not influence the
Casimir force and that the Drude or,
alternatively,  the plasma model approach is
adequate for the description of the thermal Casimir force). In this case
the exclusion of the Drude model in the experiments  \cite{30,31,42,54}
would be independently verified. This aim, however, cannot be achieved
in one experiment with magnetic materials
 because, as was mentioned in Sec.~III, the dashed lines
in Fig.\ 2(a,b) are very close to the respective
solid lines in Fig.\ 1(a,b).
This means that the role of magnetic effects in the Drude model
description nearly fully compencates differences between the theoretical
predictions using the Drude and the plasma model with magnetic effects
disregarded. Similar situation holds for Figs.\ 3 and 4. For the sake
of definiteness, we discuss below the experiments with Co plates.

Let the result of the measurement of the Casimir pressure between two Co
plates be consistent with the solid line in Fig.\ 1(b) and the dashed
line in Fig.\ 2(b). This would mean that either the metal of the bodies
is described by the Drude model and magnetic properties influence the
Casimir pressure or, alternatively,
the metal is described by the plasma model and its
magnetic properties do not influence  the Casimir interaction.
To choose between these two alternatives, a second experiment is
required. Let us consider the so-called {\it patterned} plate one half
of which is made of ferromagnetic metal (Co) and the other half of
nonmagnetic metal (Au). Let a sphere coated with  ferromagnetic metal
(Co) oscillate in the horizontal direction above different regions
of the plate. Thus, the sphere is subject to the difference Casimir
force, which can be measured using the static or dynamic techniques
\cite{70,71}. If the first alternative is correct, there is a measurable
decrease of the force magnitude when the sphere is moved from Co to Au, because
the magnitude of the free energy shown as the solid line in Fig.\ 1(a)
is larger than in Fig.\ 5(a) (remind that the force in a sphere-plate
configuration is proportional to the free energy between two parallel
plates).  If, however, the second alternative is correct, the difference
force when the sphere moves from the
Co to Au regions  takes the opposite sign.
This is because $\omega_{p, {\rm Au}}>\omega_{p,{\rm Co}}$ and the dashed line
in Fig.\ 2(a) lies lower that the dashed line in Fig.\ 6(a) (if the
 ferromagnetic and nonmagnetic metals were  selected in such a way
that their plasma frequencies would be equal,
the difference Casimir force vanishes).

Let now the results of the measurement of the Casimir pressure between two
plates coated with Co be consistent with the dashed line in Fig.\ 1(b)
and the solid line in Fig.\ 2(b). This means that either  the metal
is described by the Drude model but magnetic properties do not influence the
Casimir pressure or, alternatively,
the metal is described by the plasma model but there
is the impact of magnetic properties on the pressure magnitude.
The choice between these alternatives can be performed by the results of a
second experiment using the same patterned Co-Au plate, but with the  sphere
coated with a nonmagnetic metal (Au). If the first alternative is correct,
there is only a minor increase in the measured  force
(for about 10\% at
$a=0.5\,\mu$m) when the sphere is moved from the Co to Au
regions, as can be seen
from the solid line in Fig.\ 5(a) and respective data for Au-Au interaction
\cite{14}. If the second alternative is correct, there would be a large
increase in the measured force  (for about 20\% at
$a=0.5\,\mu$m) in the same movement (see the solid line  in Fig.\ 6(a)
and respective data for Au-Au plates \cite{14}).

Thus, the proposed measurements of the Casimir force between ferromagnetic
metals allow one not only to confirm or exclude the influence of
magnetic properties on dispersion interaction, but also to shed
additional light on the choice between different theoretical
approaches to the thermal Casimir force. Additional possibilities are
suggested by the use of the test bodies made of ferromagnetic dielectrics.
Here, in the measurement using the two plates made of ferromagnetic
dielectrics, one can determine whether the magnetic properties influence
the Casimir free energy and pressure (54\% and 36\%
relative difference,
respectively, at $a=0.5\,\mu$m, as shown in Fig.\ 9). In doing so
one does not require to make any
assumptions concerning the use of the Drude or plasma models.
Promising potentialities for the new experiments are also suggested by the
magnetic phase transition in ferromagnetic metal at Curie temperature.
According to our results, there are significant differences between the
predictions of the Drude and plasma model approaches to the thermal
Casimir force before and after the phase transition.

\section*{Acknowledgments}
 G.L.K. and V.M.M. are
grateful to the  Institute
for Theoretical Physics, Leipzig University for kind
hospitality. They are also grateful to G.\ Bimonte
for useful discussions on the early stage of the work.
This work  was supported by Deutsche Forschungsgemeinschaft,
Grants No.~GE\,696/9--1 and GE\,696/10--1.

\begin{figure*}[h]
\vspace*{-1.cm}
\centerline{
\includegraphics{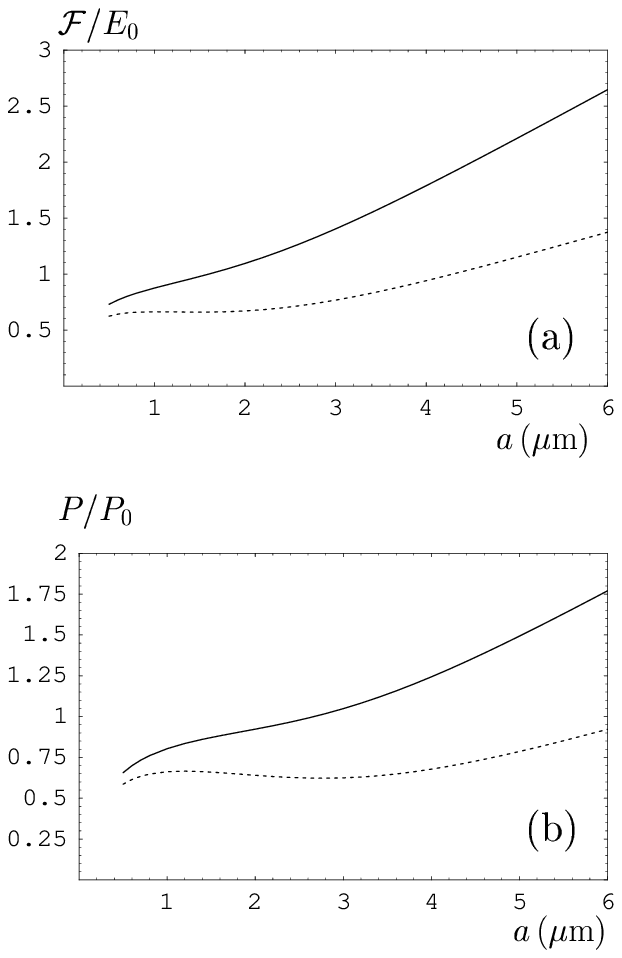}
} \vspace*{-16.3cm} \caption{The relative Casimir free energy
per unit area (a) and pressure (b) as functions of separation
in the configuration of two parallel Co plates with account of
magnetic properties (solid lines) and with magnetic
properties disregarded (dashed lines). Computations are
performed using the Drude model at $T=300\,$K.}
\end{figure*}
\begin{figure*}[h]
\vspace*{-1.cm}
\centerline{
\includegraphics{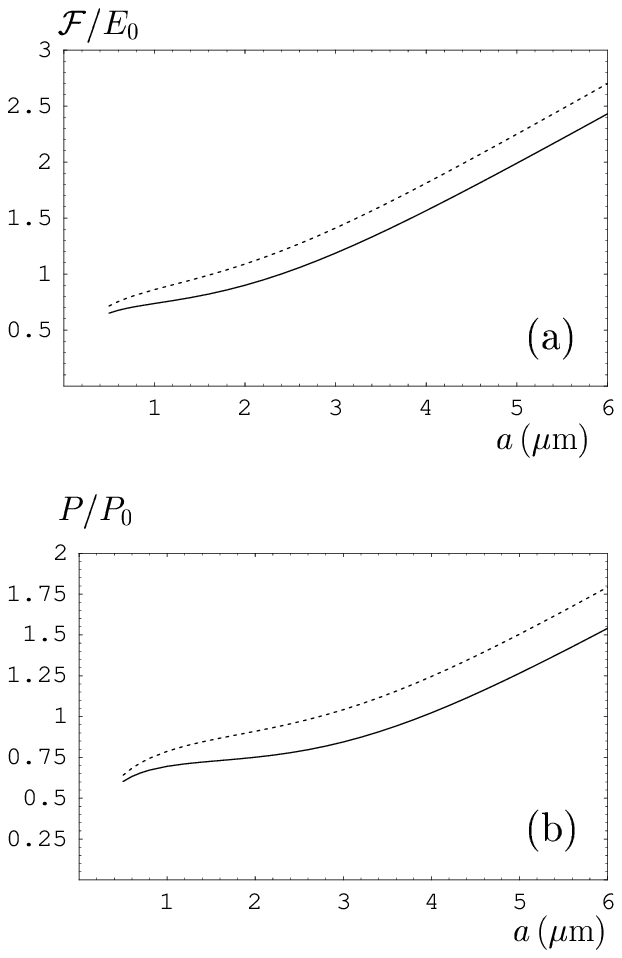}
} \vspace*{-16.3cm} \caption{The relative Casimir free energy
per unit area (a) and pressure (b) as functions of separation
in the configuration of two parallel Co plates with account of
magnetic properties (solid lines) and with magnetic
properties disregarded (dashed lines). Computations are
performed using the plasma model at $T=300\,$K.}
\end{figure*}
\begin{figure*}[h]
\vspace*{-1.cm}
\centerline{
\includegraphics{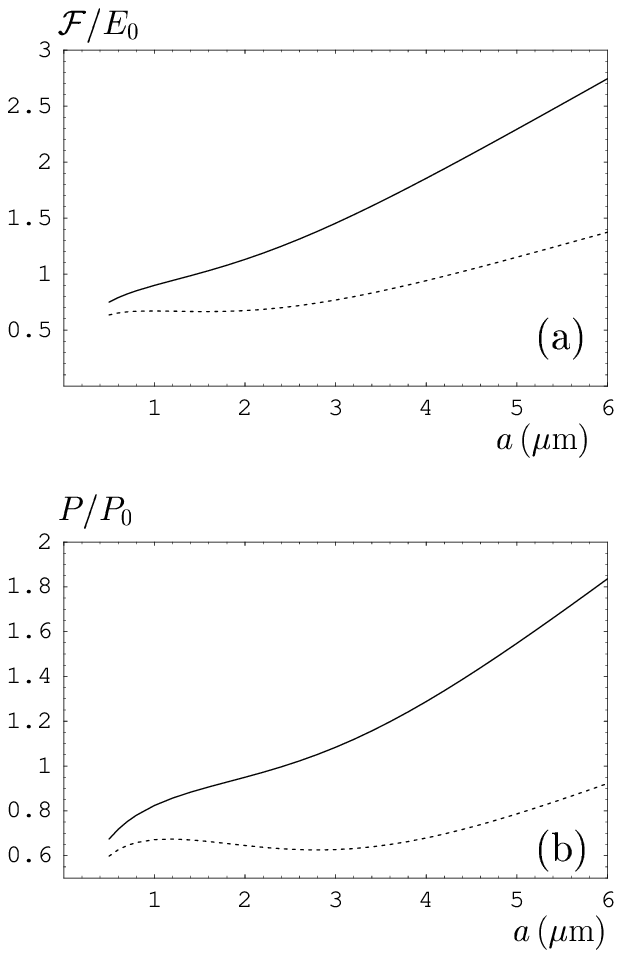}
} \vspace*{-16.3cm} \caption{The relative Casimir free energy
per unit area (a) and pressure (b) as functions of separation
in the configuration of two parallel Fe plates with account of
magnetic properties (solid lines) and with magnetic
properties disregarded (dashed lines). Computations are
performed using the Drude model at $T=300\,$K.}
\end{figure*}
\begin{figure*}[h]
\vspace*{-1.cm}
\centerline{
\includegraphics{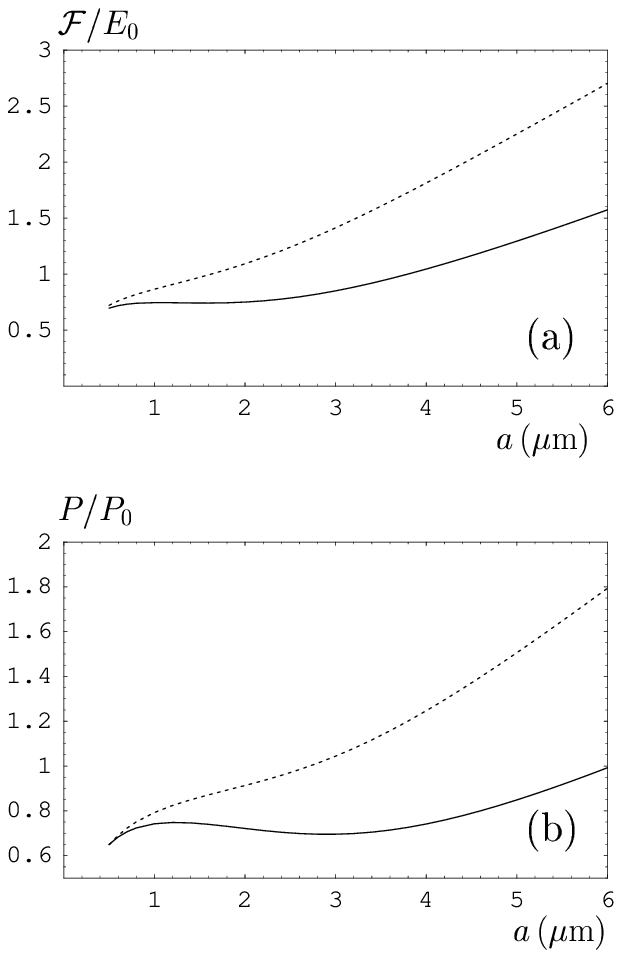}
} \vspace*{-16.3cm} \caption{The relative Casimir free energy
per unit area (a) and pressure (b) as functions of separation
in the configuration of two parallel Fe plates with account of
magnetic properties (solid lines) and with magnetic
properties disregarded (dashed lines). Computations are
performed using the plasma model at $T=300\,$K.}
\end{figure*}
\begin{figure*}[h]
\vspace*{-1.cm}
\centerline{
\includegraphics{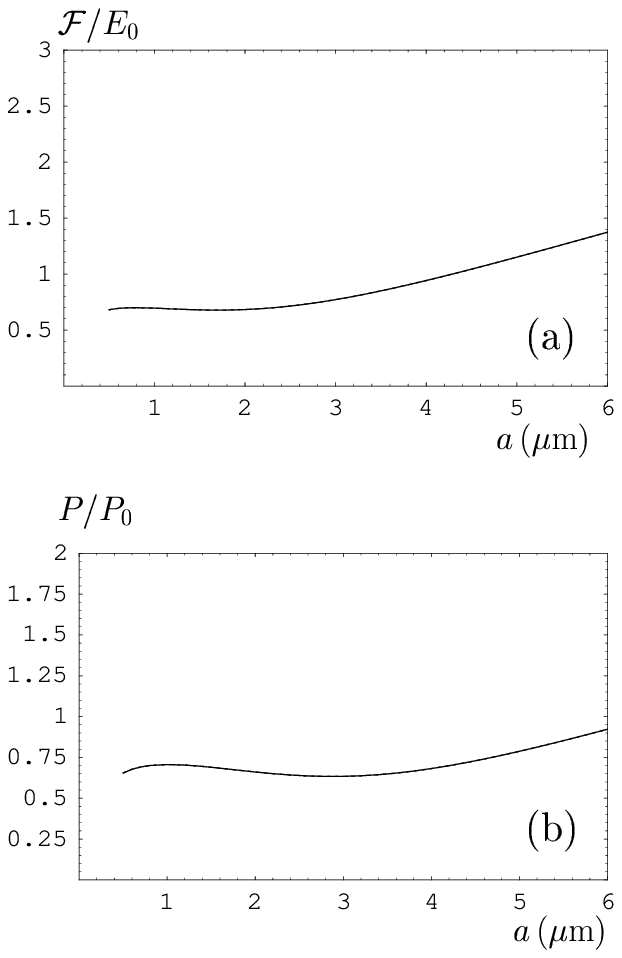}
} \vspace*{-16.3cm} \caption{The relative Casimir free energy
per unit area (a) and pressure (b) as functions of separation
in the configuration of one plate made of Co and the other
plate made of Au. Computations are
performed using the Drude model at $T=300\,$K.}
\end{figure*}
\begin{figure*}[h]
\vspace*{-1.cm}
\centerline{
\includegraphics{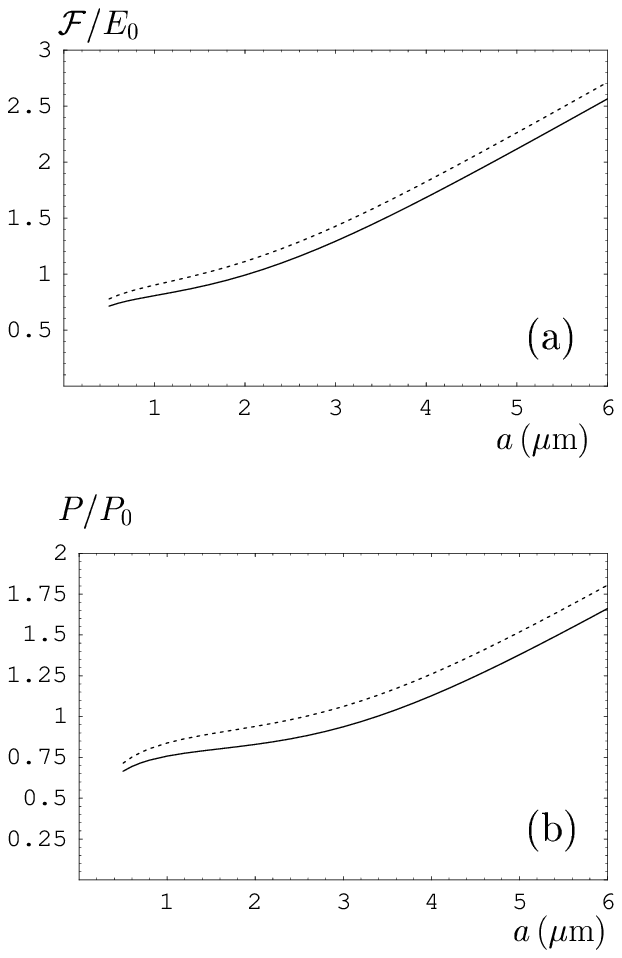}
} \vspace*{-16.3cm} \caption{The relative Casimir free energy
per unit area (a) and pressure (b) as functions of separation
in the configuration of one plate made of Co and the other
plate made of Au with account of
magnetic properties (solid lines) and with magnetic
properties disregarded (dashed lines). Computations are
performed using the plasma model  at $T=300\,$K.}
\end{figure*}
\begin{figure*}[h]
\vspace*{-1.cm}
\centerline{
\includegraphics{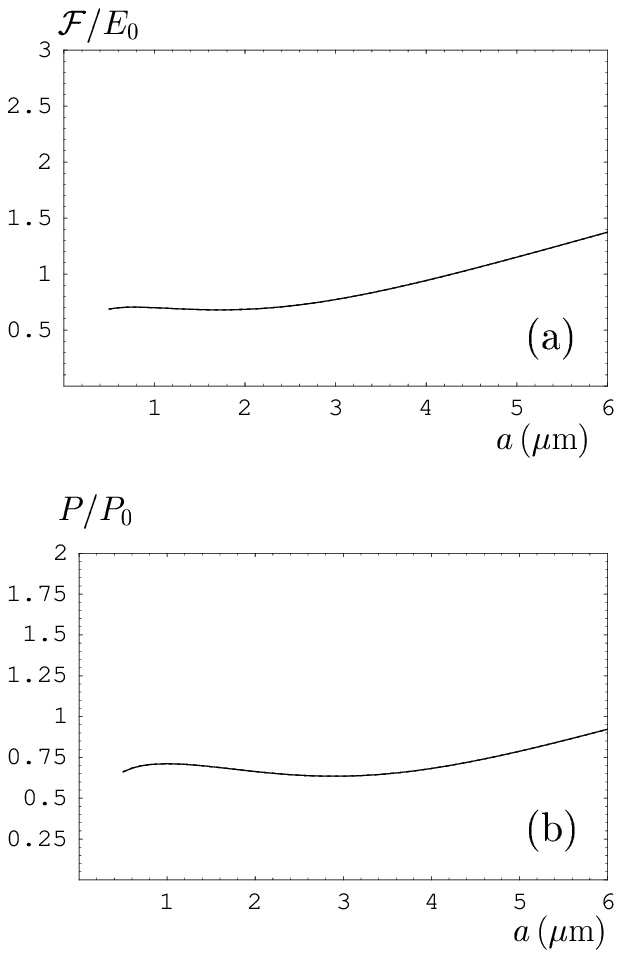}
} \vspace*{-16.3cm} \caption{The relative Casimir free energy
per unit area (a) and pressure (b) as functions of separation
in the configuration of one plate made of Fe and the other
plate made of Au. Computations are
performed using the Drude model at $T=300\,$K.}
\end{figure*}
\begin{figure*}[h]
\vspace*{-1.cm}
\centerline{
\includegraphics{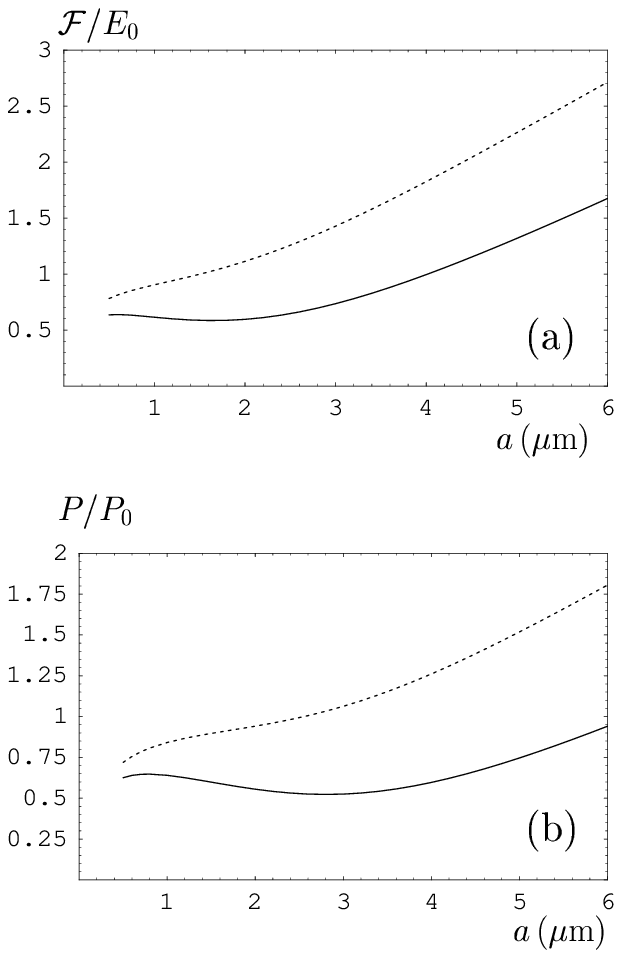}
} \vspace*{-16.3cm} \caption{The relative Casimir free energy
per unit area (a) and pressure (b) as functions of separation
in the configuration of one plate made of Fe and the other
plate made of Au with account of
magnetic properties (solid lines) and with magnetic
properties disregarded (dashed lines). Computations are
performed using the plasma model at $T=300\,$K.}
\end{figure*}
\begin{figure*}[h]
\vspace*{-1.cm}
\centerline{
\includegraphics{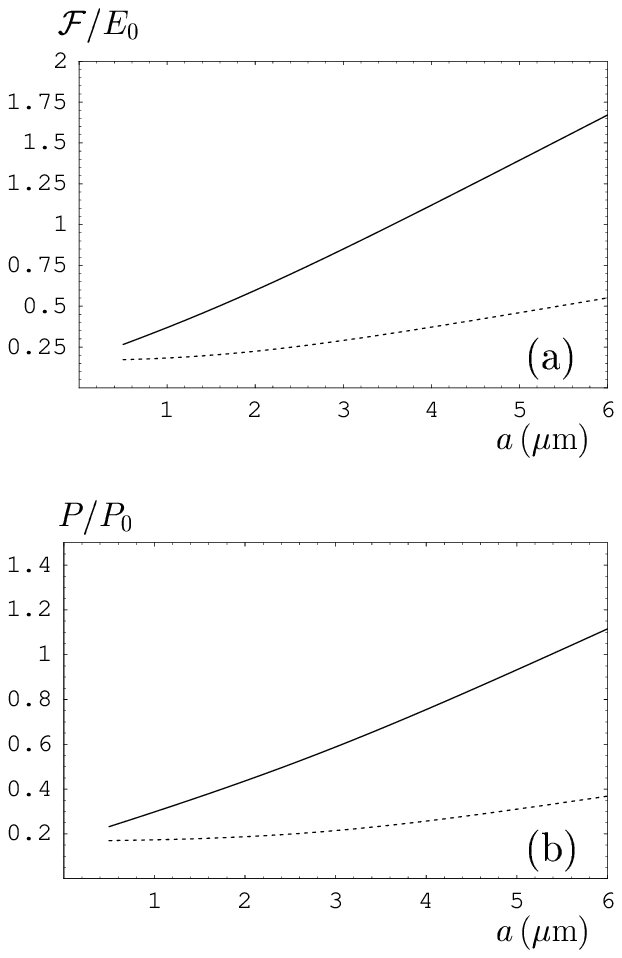}
} \vspace*{-16.3cm} \caption{The relative Casimir free energy
per unit area (a) and pressure (b) as functions of separation
 at $T=300\,$K
in the configuration of two parallel plates
made of ferromagnetic dielectric with account of
magnetic properties (solid lines) and with magnetic
properties disregarded (dashed lines).}
\end{figure*}
\begin{figure*}[h]
\vspace*{-1.cm}
\centerline{
\includegraphics{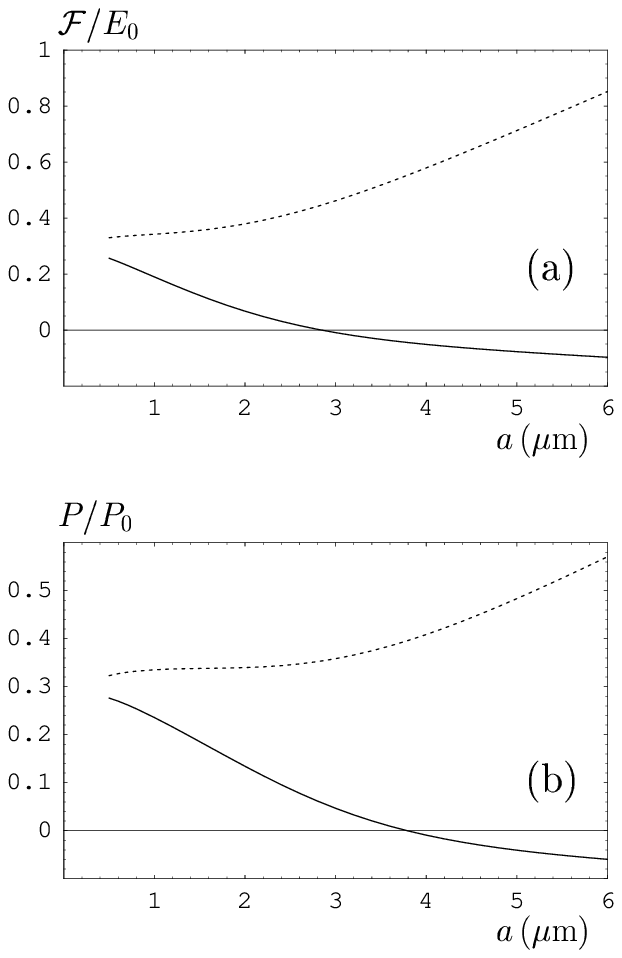}
} \vspace*{-16.3cm} \caption{The relative Casimir free energy
per unit area (a) and pressure (b) as functions of separation
 at $T=300\,$K
in the configuration of one plate made of
ferromagnetic dielectric and the other
plate made of Au with account of
magnetic properties (solid lines) and with magnetic
properties disregarded (dashed lines). Computations are
performed using the plasma model for Au.}
\end{figure*}
\begin{figure*}[h]
\vspace*{-4.cm}
\centerline{
\includegraphics{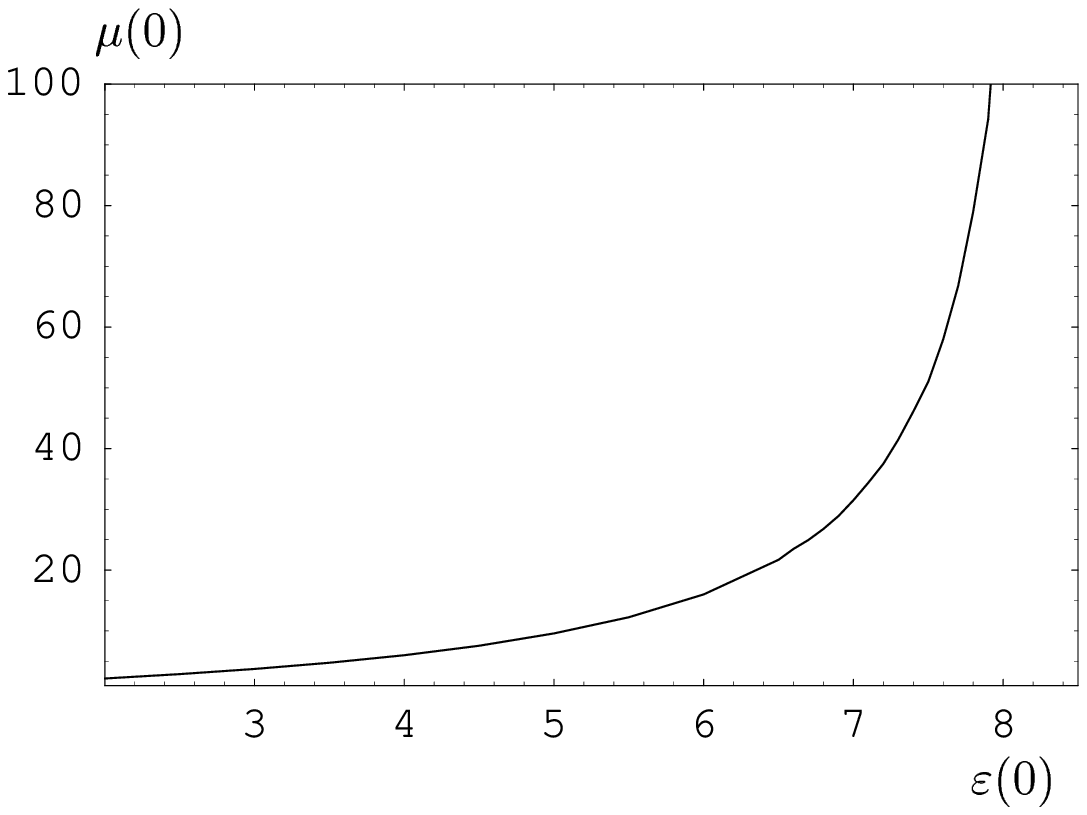}
} \vspace*{-10.3cm} \caption{The regions of the Casimir repulsion
(above the solid line) and attraction (below the solid line)
in the $[\varepsilon(0),\mu(0)]$-plane. See text for further
discussion.}
\end{figure*}
\begin{figure*}[h]
\vspace*{-4.cm}
\centerline{
\includegraphics{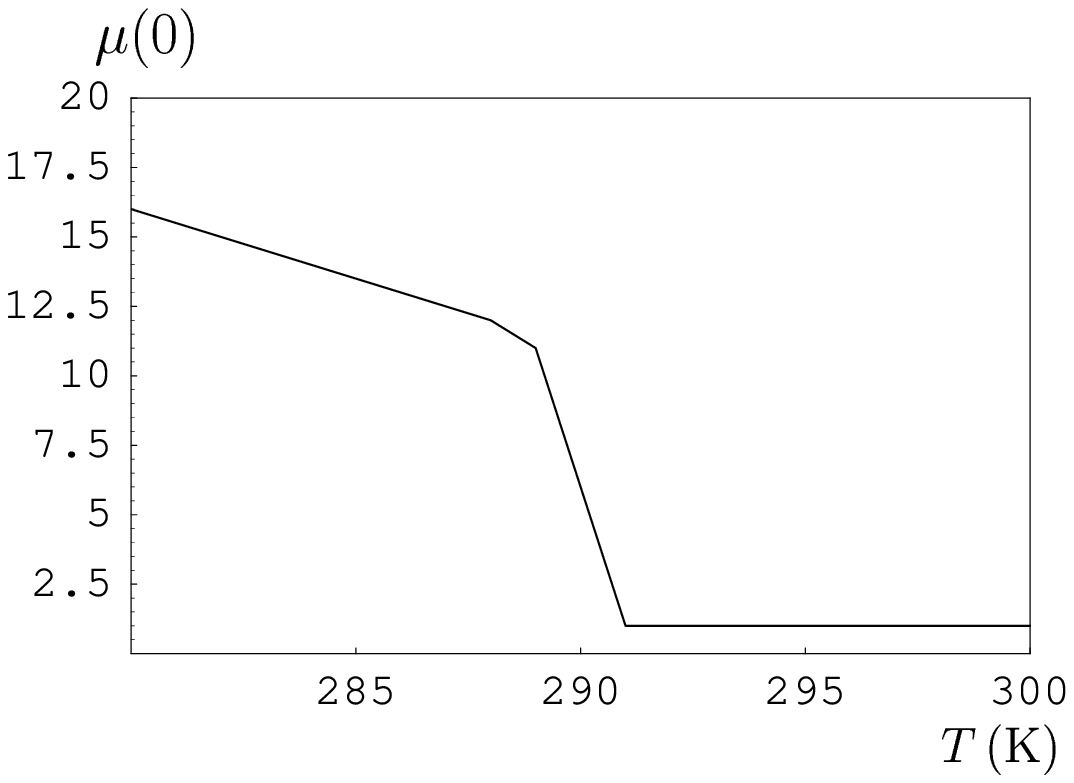}
} \vspace*{-10.3cm} \caption{The static magnetic permeability of Gd
in the magnetic phase transition as a function of temperature.}
\end{figure*}
\begin{figure*}[h]
\vspace*{-3.cm}
\centerline{
\includegraphics{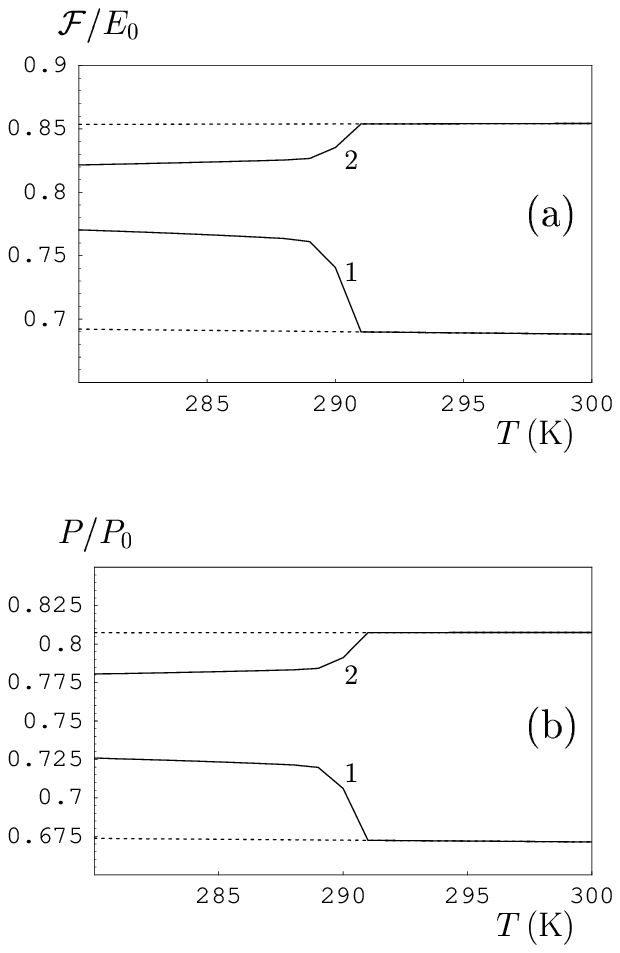}
} \vspace*{-16.3cm} \caption{The relative Casimir free energy
per unit area (a) and pressure (b) as functions of temperature
in the configuration of two parallel Gd plates at the separation
$a=0.5\,\mu$m. The solid and dashed lines take into account and
disregard the magnetic properties, respectively. The pairs
of lines marked 1 and 2 indicate the respective computational
results obtained using the Drude and plasma models.}
\end{figure*}

\end{document}